\documentclass{emulateapj}

\newcommand{\etal}{et al.~}
\def\gsim{\lower 2pt \hbox{$\, \buildrel {\scriptstyle >}\over
{\scriptstyle \sim}\,$}}
\def\lsim{\lower 2pt \hbox{$\, \buildrel {\scriptstyle <}\over
{\scriptstyle \sim}\,$}}

\def\ciao{{\sl CIAO~}}

\def\chandra{{\sl Chandra~}}
\def\oix{O~{\scriptsize IX}}
\def\oviii{O~{\scriptsize VIII}}
\def\ovii{O~{\scriptsize VII}}
\def\ovi{O~{\scriptsize VI}}
\def\oi{O~{\scriptsize I}}
\def\oii{O~{\scriptsize II}}
\def\oiv{O~{\scriptsize IV}}
\def\oiii{O~{\scriptsize III}}
\def\neix{Ne~{\scriptsize IX}}
\def\nex{Ne~{\scriptsize X}}

\def\aliii{Al~{\scriptsize III} }

\def\source{4U~1820--303}

\shortauthors{Yao \& Wang}
\shorttitle{Oxygen and Neon Abundances in the Multiple Phase ISM}

\begin{document}
\slugcomment{\em Accepted for publication in the  Astrophysical Journal}

\title{X-ray Absorption Spectroscopy of the Multi-phase Interstellar 
Medium: Oxygen and Neon Abundances}

\author{Yangsen Yao\altaffilmark{1,2} and Q. Daniel. Wang\altaffilmark{2,3}}
\altaffiltext{1}{MIT Kavli Institute for Astrophysics and Space Research, 70 Vassar Street, Cambridge, MA 02139; yaoys@space.mit.edu}
\altaffiltext{2}{Department of Astronomy, University of Massachusetts, 
  Amherst, MA 01003; wqd@astro.umass.edu}
\altaffiltext{3}{Institute for Advanced Study, Einstein Drive, Princeton, NJ 08540}

\begin{abstract}

X-ray absorption spectroscopy provides a potentially powerful
tool in determining the metal abundances in various phases 
of the interstellar medium (ISM).
We present a case study of the sight line toward \source\ 
(Galactic coordinates $l, b=2^\circ.79, -7^\circ.91$ and distance = 7.6 kpc), based on 
{\sl Chandra} Grating observations. The detection of \oi, \oii, \oiii, \ovii, 
\oviii, and \neix\ K$\alpha$ absorption lines allows us to measure 
the atomic column densities of the neutral, warm
ionized, and hot phases of the ISM
through much of the Galactic disk. The hot phase of the ISM 
accounts for about 6\% of the total oxygen column density
$\sim8\times 10^{17} {\rm~cm^{-2}}$ along the sight line,
with the remainder about evenly divided between the neutral and warm
ionized phases. By comparing these measurements 
with the 21 cm hydrogen emission and with the pulsar dispersion measure 
along the same sight line, we estimate the mean oxygen abundances in the 
neutral and total ionized phases as 0.3(0.2, 0.6) and 2.2(1.1, 3.5)
in units of \citet{and89} solar value (90\% confidence intervals).
This significant oxygen abundance difference is apparently a result
of molecule/dust grain destruction and 
recent metal enrichment in the warm ionized and hot phases. 
We also measure the column density of neon
from its absorption edge and obtain the 
Ne/O ratio of the neutral plus warm ionized gas as
2.1(1.3, 3.5) solar. Accounting for the expected
oxygen contained in molecules and dust grains would reduce the Ne/O ratio
by a factor of $\sim 1.5$. 
From a joint-analysis of the \ovii, \oviii, and \neix\ lines, we
obtain the Ne/O abundance ratio of the hot phase as 1.4(0.9, 2.1) 
solar, which is not sensitive to the exact 
 temperature distribution assumed in the absorption line modeling.
These comparable ISM Ne/O ratios for the hot and cooler gas
are thus considerably less than the value ($2.85 \pm 0.07$; 1$\sigma$)
recently inferred from corona emission of solar-like stars 
(Drake \& Testa 2005).

\end{abstract}
\keywords{ISM: abundances --- X-rays: ISM --- X-rays: individual (\source)}

\section{Introduction \label{sec:intro}}

The measurement of the ISM metal abundances plays a key role in 
our understanding of the universe. 
Although metals were all produced in stars, the ISM is
the depository of stellar feedback (e.g., via supernovae). 
The metal abundances of the ISM 
thus give a unique measure of the integrated stellar feedback. The ISM is
also a reservoir from which stars formed; the similarity
or dissimilarity of the abundances in the ISM and stars  provides 
important constraints on the
physical processes involved in star formation and evolution (e.g.,
metal diffusion and settling). Of course, the abundances are also important in
the study of various physical and chemical processes of the ISM itself 
(e.g., heating and cooling as well as dust grain formation). 
Furthermore, the interpretation of 
existing observational data often relies heavily on the assumption of the 
abundances in various phases of the ISM. Examples are the modeling of hot 
plasma spectra and the correction for ISM absorption to infer 
the intrinsic spectra of X-ray sources. Therefore, the measurement
of the abundances represents a major task of astrophysics.

But a complete accounting of the metal abundances in the ISM has been 
a challenge. The ISM is a heterogeneous ensemble of 
various forms (atomic, molecular, and solid dust grain)
and various states (phases): cold ($\lsim 100$ K),
warm ($\sim8000$ K; including warm neutral and warm photo-ionized), and 
hot ($\sim10^{6}$ K; collisionally ionized) 
(e.g., Ferri\'ere 2001 and references 
therein). For easy of reference in the present work, we will not distinguish 
the cold and
warm neutral phases and will call their combination as the ``cold'' phase. 
The ``warm'' phase will, in stead, refer only to the above warm ionized phase;
the total ``ionized'' phase therefore refers to a combination of warm and hot 
phases. Also these phases are used for the {\sl atomic} gas form only, whereas 
molecule and solid dust grains are denoted as the
``compound'' form without any phase separation.
The abundances of individual elements
could be significantly different in these forms and phases, because of the
varying level of depletion and molecule/grain destruction as well as 
recent chemical enrichment from stellar feedback. The
traditional methods for abundance measurement in the optical and ultraviolet 
(UV) are sensitive mainly to the cold and warm phases. 
The reliability of such measurements strongly depends on how well the
physical and ionization conditions are modeled (e.g., Savage 
\& Sembach 1996 and  references therein).

Stellar abundance measurements also suffer large uncertainties, especially
for solar-like  stars.
Recently, solar abundances of light elements such as C, N, O and Ne 
have been revised downward by 25-45 per cent \citep{asp05} 
from the values of Anders \& Grevesse (1989; AG89 hereafter).
The revised values reasonably agree with 
UV and X-ray measurements of the O abundance in the cold and warm ISM phases
(e.g., Sofia \& Meyer 2001; Takei \etal 2002; Andr\'e \etal 2003; this work). 
However, this revision 
has broken the abundance accordance with helioseismological measurements 
and theoretical solar model predictions (e.g., Bahcall et al. 2005a).
It is argued that the Ne abundance in the Sun is poorly determined
and that if the Ne/O ratio is in fact substantially larger (by a factor 
of $\gtrsim 2.5$) the models can then be brought back into agreement with 
helioseismological measurements  (e.g., Bahcall et al. 2005b).
Based on the spectroscopy of the \oviii, \neix, and \nex\
emission lines in the 
\chandra HETG-ACIS spectra of nearby solar-like stars, \citet{drake05}
estimate that the averaged Ne/O number ratio is $0.41\pm0.01$ (1 $\sigma$), 
or $2.85\pm0.07$ of the AG89 value. However, 
two latest independent studies suggest
no such high Ne/O ratio at the solar surface \citep{sch05, young05}.

X-ray absorption spectroscopy can, in principle, be the ideal tool 
to measure the metal abundances of the ISM, in essentially all 
forms and phases. 
The X-ray bandpass contains almost all the K- and L-transitions of different 
charge states of the abundant elements from carbon to iron. 
X-ray spectroscopy, less affected by extinction, 
also probes larger column densities than is possible in the optical
and UV, and is thus especially useful for measuring
the general ISM through much of the Galactic disk. With the still limited
spectral resolution and sensitivity of existing X-ray telescopes, however, 
only the most abundant elements (like oxygen and neon) can
produce significant absorption features in the spectra of bright background
sources such as AGNs and X-ray binaries. With the grating instruments 
aboard \chandra and {\sl XMM-Newton}, several groups have indeed studied
the abundances of X-ray-absorbing gas, based chiefly
on measurements of absorption edges, which are contributed by the cold and warm
phases in both atomic and compound forms
\citep{pae01, sch02b, tak02, jue03, jue05, 
cunn04, ued05}. Most of these studies could only get relative metal 
abundances (e.g., Ne/O). But \citet{tak02} and \citet{cunn04} were also 
able to estimate the absolute 
abundance (e.g., O/H) for the sight lines to Cyg X--2 and X Persei, 
by using emission line data (e.g, 21 cm)
and UV absorption lines. Some success has been achieved even in
distinguishing the contributions from the atomic and compound forms
through calibrating the still uncertain
wavelengths and cross-sections of the absorption edges \citep{tak02}.

More recently, Yao \& Wang (2005; hereafter Paper I) and 
Wang \etal (2005; hereafter Paper II) have demonstrated 
the possibility of measuring the abundances of the hot ISM phase. By jointly
fitting various highly-ionized species, significantly detected or not,
one can simultaneously determine multiple parameters of the modeled 
absorbers such as the gas temperature, velocity dispersion, 
and element abundance ratio. 
Even with the limited counting statistics of the LMC X-3 observation, 
\citet{wang05} are able to obtain a meaningful lower limit to the 
Ne/O ratio, which is about the solar value of AG89.

We also note that the detection of \ion{O}{1}, \ion{O}{2}, and \ion{O}{3} 
absorption lines allow for direct measurements of the cold and warm
oxygen column densities. These lines have large oscillator 
strengths relative to the expected velocity dispersions of the ISM phases;
there is little flat part in the curve-of-growth (CoG) of such 
a line.  Therefore, one can reliably measure the column densities 
in the typical square-root ranges of the CoG.

Here we capitalize on the high-quality detections of 
multiple absorption lines in the \chandra\ spectra of \source\ (Futamoto \etal 
2004; Paper I) to estimate the abundances in the multi-phase ISM of the 
Galaxy. Throughout this paper, we use the AG89 solar abundances as a
convenient reference; the number ratios of O/H and Ne/O are
8.5$\times10^{-4}$ and 0.144; in comparison, the 
recently revised solar values are $4.6\times10^{-4}$ and 
0.151, respectively (Asplund, Grevesse, \& Sauval 2005; further discussion in \S~5). 
We also assume a collisional ionization equilibrium 
(CIE) in the hot intervening X-ray-absorbing gas; this should be quite a 
good approximation for the oxygen and neon in typical hot ISM conditions
\citep{sut93}. The quoted  
parameter errors are all at 90\% confidence levels
unless otherwise specified.

\section{\source\ and \chandra Data Calibration \label{sec:obs} }

\source\ is an exceptionally good target for the X-ray absorption
spectroscopy of the ISM. First, this low mass X-ray binary (LMXB) 
 is super compact, with an orbital period of only 685 s \citep{ste87}; 
the large ionization parameter in the immediate vicinity of the binary
greatly reduces the chance for a significant
local contribution to the observed  absorption lines,
consistent with their constancy, moderate ionization state, and
insignificant width/shift (Futamoto \etal 2004;
Paper I). Second, the binary resides in the globular 
cluster NGC~6624 ($l$, $b$ = $2^\circ.79$, -$7^\circ.91$, distance $= 
7.6\pm0.4$ kpc; Kuulkers \etal 2003) near the Galactic center 
(distance $ = 7.62\pm0.32$ kpc; Eisenhauer \etal 2005). Therefore, the 
pathlength of the sight line is well determined and samples the entire 
inner Galactic disk radially to a height of 
$\sim 1$ kpc off the Galactic plane. Third, the globular cluster also
contains two radio pulsars, PSR~1820-30A and 
PSR~1820-30B, with essentially the same dispersion measure (DM)
of 87 cm$^{-3}$~pc or a total free electron column density of 
${\rm N_e = 2.7 \times10^{20}~cm^{-2}}$ \citep{big94}, 
very useful for an absolute abundance measurement of ionized gas.

\chandra\ observed \source\ three times: one observation used the low-energy
transmission grating with the high resolution camera 
(LTEG-HRC; Brinkman \etal 2000) at the focal
plane for 15 ks, and the other two used the high-energy transmission grating 
with the advanced CCD imaging spectrometer 
(HETG-ACIS; Canizares \etal 2005) 
for 9.7 and 10.9 ks, respectively.
The LETG-HRC observation, particularly sensitive to \ovii,
 has been reported by \citet{fut04}, while the 
HETG-ACIS observations, sensitive to \neix, have 
been studied together with other targets by Juett, Schulz, \& 
Chakrabarty (2004) and in Paper I. 
These studies have focused 
on either the global distribution of the intervening hot gas 
(Futamoto \etal 2004; Paper I) or on the oxygen absorption edge structure 
\citep{jue04}.
Here we combine all these observations of \source, improving both the
energy coverage and the counting statistics, to measure
oxygen and neon abundances in the cold, warm, and hot phases.

We reprocess the three observations, using \ciao~3.2.1 and the calibration 
database CALDB~3.1.0\footnote{http://cxc.harvard.edu/ciao/download}. 
We use \ciao\ thread {\sl tgextract} to extract the
source spectra. We rebin the MEG spectra to a bin size of 0.0125 \AA\
so that they can be easily  combined with the LETG spectrum.
For each observation, we calculate the response matrices files (RMFs) and 
effective area files (ARFs) for the 
positive- and negative-grating arms by running scripts 
{\sl mkgrmf} and {\sl mkgarf}. For the HETG-ACIS observations,
the effective area of the MEG is $\sim5$ times larger than that of the 
high energy grating (HEG) in the wavelength band where our diagnostic 
absorption lines are located (e.g., longward of 13\AA; Table~\ref{tab:gauss}). 
The effective area of the 1st orders is also at least one order 
of magnitude larger than that
of the higher orders ($\ge2$). Therefore, we only utilize the 1st order 
MEG spectra in this study. For the LETG-HRC observation, the detector has 
little intrinsic energy resolution, thus one needs to account for
not only the local line spread but also the global spectral
overlapping of different orders. We first calculate the RMFs and ARFs from
the 1st to 6th grating orders [assuming the higher ($>6$) order contribution to
the spectrum in our interested wavelength range is negligible; see below]
and then add these RMFs and ARFs pairs 
to form order-combined response files (RSPs; see 
Nicastro \etal 2005 and Paper II for more details).
To improve the counting statistics, we further co-add the positive- and
negative-grating MEG spectra 
and  combine them with the LETG-HRC spectrum to form 
the final LETG+MEG spectrum. The corresponding RMFs and ARFs
(RSPs) are averaged, channel by channel, with a globally modeled spectrum 
as the weighting function (similar to what is described below
for the combined spectrum). We conduct various 
consistency checks for key results, based on the spectra from individual 
observations.

To account for the spectral order overlap, we model our combined
spectrum over a broad range of 2-24.5 \AA. As in previous studies
(Nicastro \etal 2005 and Paper II),
we fit the continuum with a combination of models:
a blackbody (BB) and a broken power-law (BKNPL), plus ten broad 
($\sigma>1000$ km~s$^{-1}$) Gaussians (3 negative and 7 positive). 
These broad Gaussian features are used to compensate for both the deviation
of the data from the simple BB+BKNPL combination and the 
calibration uncertainty in various gap/node regions. In addition,
a foreground absorption with solar metal abundances 
is also applied. This continuum model
gives a fit with $\chi^2/d.o.f$=2448/1784. The best-fit 
hydrogen column density is $N_H = 2.0\times10^{21}$ cm$^{-2}$,
consistent with the values obtained from the optical extinction 
and \ion{H}{1} 21 cm emission measurements (see \S~4). This global fit is
sufficiently accurate for our order overlap modeling. 
The HRC-LETG higher order 
contributions are relevant at wavelengths longward of 10 \AA\ and 
are $\sim7\%$ and 40\% at 15 and 20 \AA, respectively. Our subsequent 
spectroscopy is based primarily on local spectral data around individual
absorption edge/lines considered.
The spectral analysis uses the X-ray spectral analysis
software package XSPEC (version 11.3.1). 

\section{Analysis and Results \label{sec:results} }

The detected \oi, \oii, \oiii, \ovii, \oviii, and 
\neix\ K$\alpha$ absorption lines (see also Juett \etal 2004; Futamoto \etal 2004; 
Paper I) are clearly visible in our combined spectrum. 
We use six narrow negative Gaussian profiles to characterize these lines,
which improves the spectral fit ($\chi^2/d.o.f$=2042/1768). 
Table~\ref{tab:gauss} summarizes this Gaussian characterization of the 
absorption lines, including the velocity shifts ($cz$) relative to
the adopted rest-frame wavelengths ($\lambda$) and
the equivalent widths (EWs) as well as the significance of 
each line ($\Delta \chi^2$). We note that the \oii\ K$\alpha$ line
\citep{jue03} may be
significantly contaminated by a weak oxygen compound line at
$\sim23.35$ \AA\ \citep{tak02}. But the wavelength and oscillator strength
of the line (or complex) are very uncertain, and depend on the exact 
composition of the compound. All the observed lines are unresolved; 
the instrument resolution gives an upper limit  ${\rm \sim 500~km~s^{-1}}$
to the velocity dispersion ($\sqrt{2}$ times the Gaussian width)
of the \ovii~K$\alpha$ line.
We also obtain an upper limit to the EW of the \ovii~K$\beta$ line by
fixing its centroid and linking its velocity dispersion with that of 
\ovii~K$\alpha$ (Table~\ref{tab:gauss}). This Gaussian characterization 
of the lines is consistent with previous studies
(Futamoto \etal 2004; Juett \etal 2004; Paper I), although
our present results have smaller 
uncertainties because of the improved counting statistics in the 
LETG+MEG spectrum. Since all the lines, except for 
\ovii~K$\beta$, are saturated, the EWs
(presented here only for ease of comparison with similar analyses) 
could be strongly biased due to the inaccuracy of the Gaussian 
characterization 
(Paper I) and should thus be interpreted with caution.

\begin{deluxetable*}{lccccc}
\tablewidth{0pt}
\tablecaption{Line parameters and Gaussian fit results \label{tab:gauss}}
\tablehead{
       & $\lambda$ & &   & $cz$         &  EW   \\
  line & (\AA)     &$f_{ij}$ & $\Delta\chi^2$ & (km s$^{-1}$) & (m\AA) 
}                                      
\startdata                    
\oi~K$\alpha$       & 23.508 & 0.431& 205 &  4(-92, 124)    & 83(66, 141) \\
\oii~K$\alpha$      & 23.352 & 0.424& 36  &  1(-287, 106)   & 86(52, 173) \\
\oiii~K$\alpha$     & 23.106 & 0.141& 28  & 251(-47, 550)   & 74(41, 109) \\
\hline
\ovii~K$\alpha$     & 21.602 & 0.696& 49  &  204(89, 328)   & 40(30, 53) \\
\ovii~K$\beta$      & 18.627 & 0.146& 2   & fixed	   & $<$ 10 \\
\oviii~K$\alpha$    & 18.967 & 0.277& 34  & -48(-215, 119) & 19(13, 25) \\
\neix~K$\alpha$     & 13.448 & 0.657& 52  & -35(-213, 68)  & 9(7, 11) \\
\enddata
\tablecomments{
  The rest-frame wavelength ($\lambda$) values of the \oi, \oii, and \oiii\ K$\alpha$ 
  absorption lines are from averages over multiple detections;
absolute theoretical values are still quite uncertain
  (see Juett \etal 2004; Paper II). The transition oscillator 
  strength $f_{ij}$ values of these lines are from \citet{gar05}. 
  The $\lambda$ and $f_{ij}$ values of the other lines are as in Paper I. 
  $\Delta\chi^2$ is the $\chi^2$ reduction when the corresponding 
  line is included in the spectral fit. }
\end{deluxetable*}

Our analyses and results, presented in the following sections, 
are based on the recently 
constructed absorption line model {\sl absline}, as detailed in Paper I. 
This model adopts the accurate {\sl Voigt} function as the
line profile, accounting for line saturation automatically and allowing
for a joint-fit to multiple absorption lines. The basic model parameters are
the line centroid $E_l$, velocity dispersion $b_v$, 
and reference ionic column density $N{\rm_X}$ (e.g., $X$=\ovii\ or H). In a fit to multiple lines, other parameters can be included, such as
the temperature $T$ and metal abundance(s) 
$Z_X$. Such a fit can be much more powerful than a simple CoG analysis: e.g., 
parameters such as the velocity dispersion can be measured from relative line
saturations, and statistical uncertainties in multiple parameters
can be easily determined. 
Paper II includes a detailed comparison of the {\sl absline} model 
with the CoG analysis, which is 
more illustrative but less powerful in the present context.

\subsection{Cold and Warm Phases}

\begin{figure*}
  \centerline{
    \vbox{
      \hbox{
	\includegraphics[width=0.45\textwidth]{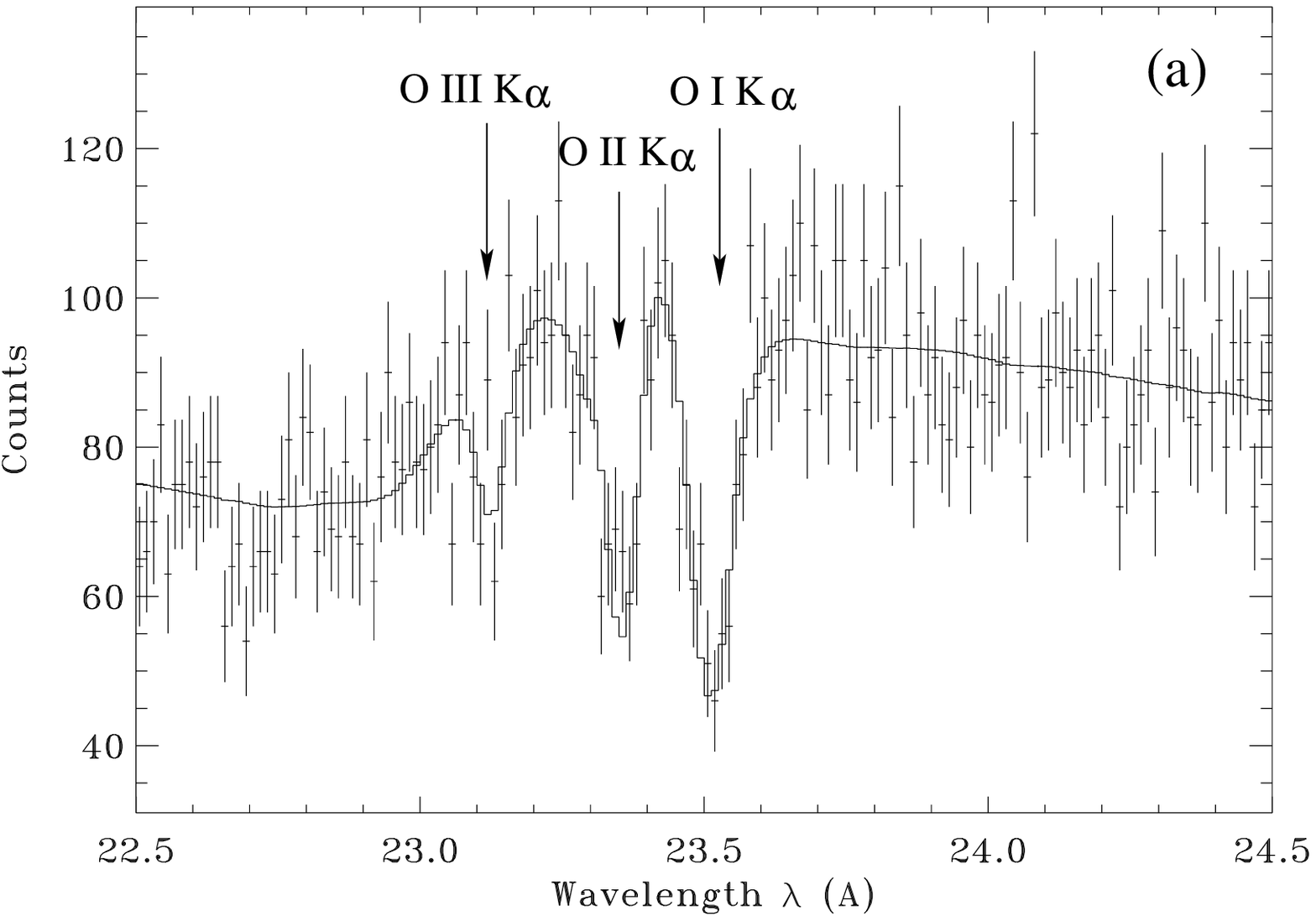}
	\includegraphics[width=0.45\textwidth]{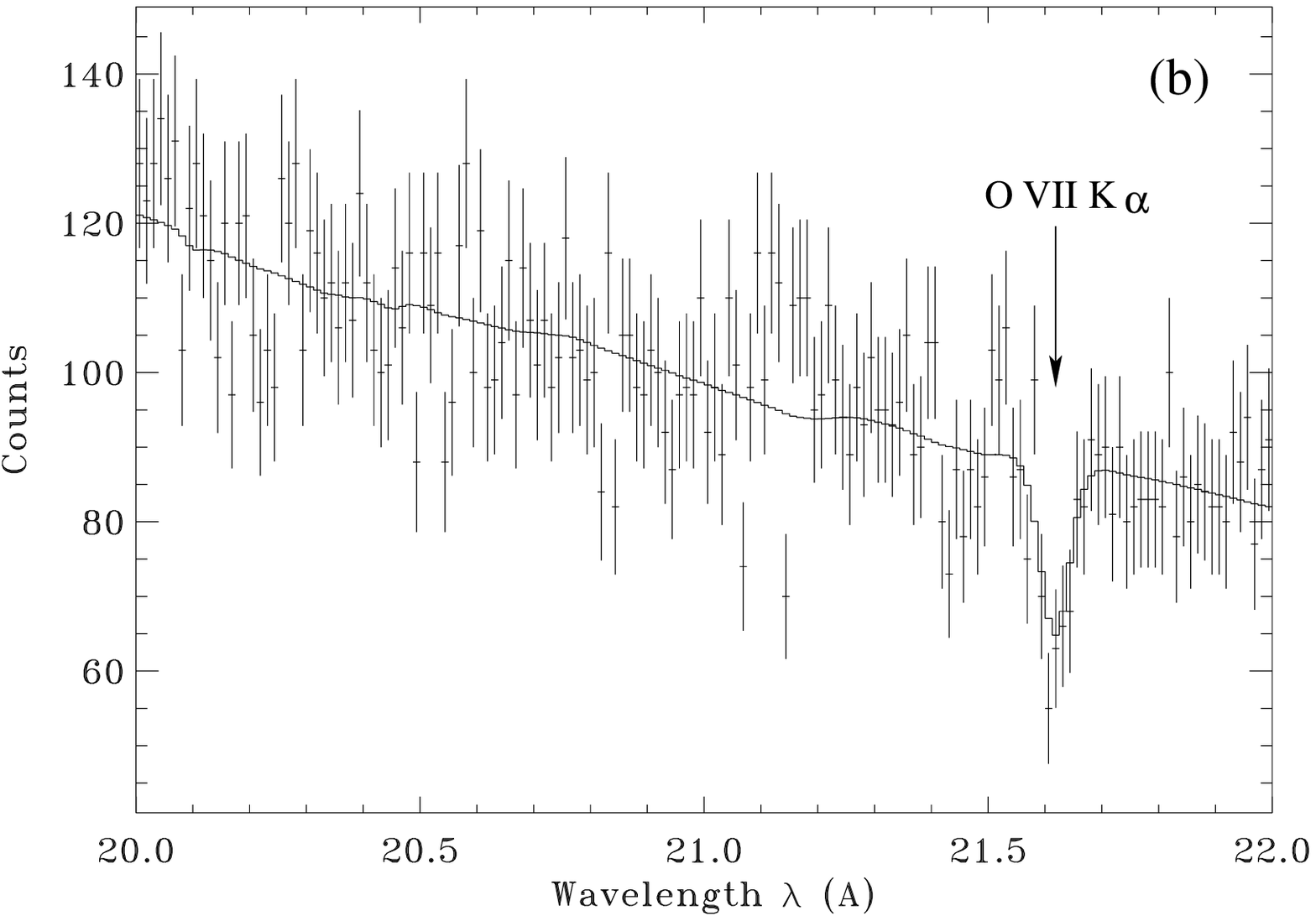}}
      \hbox{
	\includegraphics[width=0.45\textwidth]{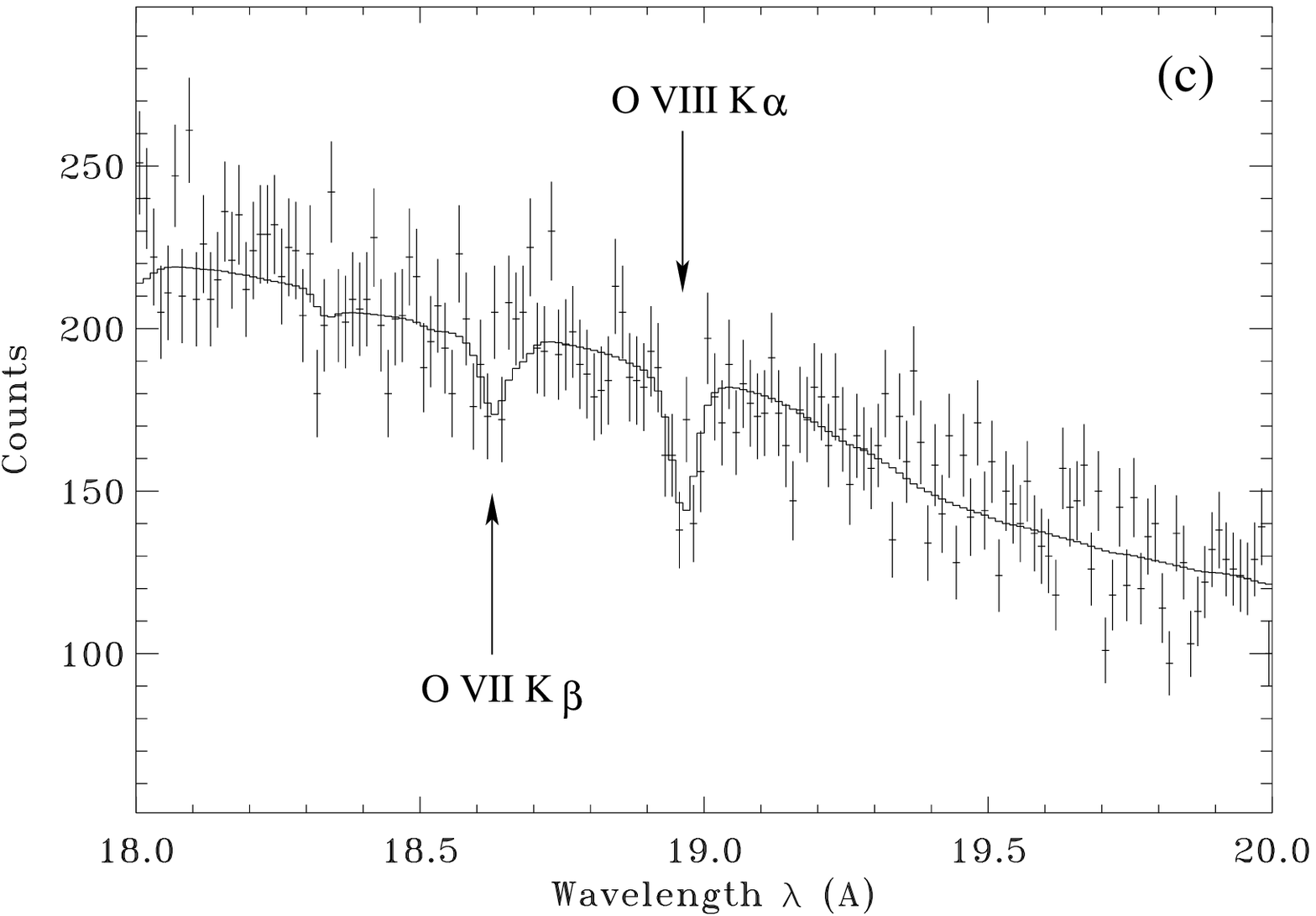}
	\includegraphics[width=0.45\textwidth]{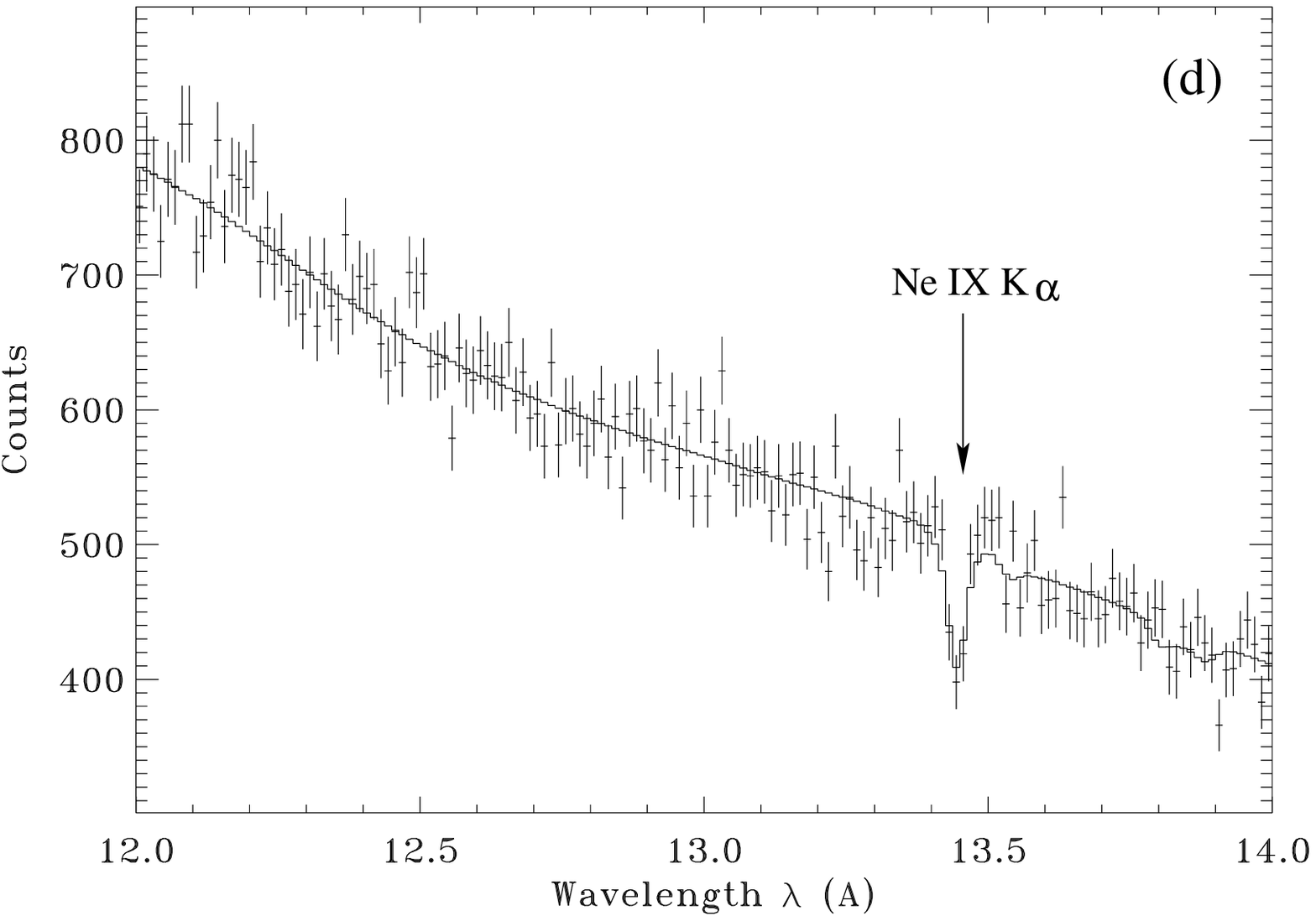}}
    }
  }
  \caption{The oxygen and neon absorption lines in the LETG+MEG
    spectrum (with a bin size of 0.0125 \AA), together with the best-fit 
    {\it absline} model plus the continuum.
    The ARF and RMF have been applied to the model. 
    The local $\chi^2/d.o.f.$ are 
    156/130 (a), 146/154 (b), 203/153 (c), and 177/153 (d). The wavelength
    is marked for the 1st order, although the spectra are also contributed
    by higher orders. The \ion{O}{7} K$\beta$ (c), although not significantly
    detected, is included in the fit. See text for detail.
    \label{fig:OI} }
\end{figure*}

We fit the \oi, \oii, and \oiii~K$\alpha$ lines,  individually, with 
the {\sl absline} model [Fig.~\ref{fig:OI} (a)]. 
We assume a velocity dispersion $b_v = 62$ km~s$^{-1}$,
inferred from the \aliii\ absorption line in the high-resolution 
far-UV spectrum of HD 163522, a star in an adjacent 
direction of ($l$, $b$)=(350$^\circ$, -9$^\circ$) and at a distance of
$\sim9$ kpc \citep{sav90}. 
The fitted column densities are 
log[$N{\rm _{OI}(cm^{-2})}$] = 17.6(17.3, 17.9), 
log[$N{\rm _{OII}(cm^{-2})}$] = 17.4(16.9, 17.6), and 
log[$N{\rm _{OIII}(cm^{-2})}$] = 17.0(16.5, 17.5). These values 
are not very sensitive to the exact choice
of the assumed $b_v$ value, because these X-ray absorption lines  
are in the saturated square-root range of the CoG (\S 1); 
a factor of two variation in $b_v$, for example, would cause 
a change $\lesssim 20\%$ in these column densities, which is
 smaller than the statistical uncertainties.
For \oi, in particular, we find that the column density changes by
$\lsim10\%$ when $b_v$ changes from 62 to 0 km~s$^{-1}$. 
Systematically, however, $N{\rm _{OII}}$ might be over-estimated
because the potential confusion with compound oxygen absorption
(e.g., Takei \etal 2002; further discussion in \S 4).

These column density estimates, based on absorption lines, may be compared 
with the absorption edge results of \citet{jue04} from the same source. 
They measured the column density of oxygen only in the atomic form, although 
the atomic edges are embedded within similar structures produced by 
oxygen in the compound form.
The compound edge structures are at lower energies and are less well
defined than the atomic ones because of such uncertain factors as grain 
and molecule composition and confusion with various absorption lines
\citep{jue04, tak02}. In addition to atomic edges, \citet{jue04} also 
focused on the identification and calibration of the K$\alpha$ lines
from \oi, \oii, and \oiii. Their analysis was based on 
HETG observations of seven
X-ray binaries, including \source. Toward this source, they first
obtained an oxygen column density of 1.31(1.17, 1.51)$\times10^{18}$ 
cm$^{-2}$ from the edge measurement and then predicted an EW
of the \oi~K$\alpha$ absorption line. This EW is, however, significantly
higher than the observed value. Similar offsets are also 
apparent for most of the other 
sight lines. They tentatively attributed this difference
to their use of the Gaussian line function, instead of the more accurate 
Voigt profile. While the latter profile is used in our analysis, the
offset for \source\ remains; i.e., our measured \oi\ column density 
${\rm 4.0(2.0, 7.9) \times10^{17}~cm^{-2}}$ is 
substantially smaller than the oxygen column density from the edge
measurement. In fact, an offset {\sl is} expected because 
the oxygen absorption edge includes contributions not only from \oi,
as is traced by its K$\alpha$ line, but also from the oxygen in the compound 
form and from \oii~and \oiii~in the warm phase of the 
ISM. The sum of our measured column densities of \oi, \oii, and \oiii\ gives
N(\oi+\oii+\oiii)=${\rm 0.75(0.48, 1.22)\times10^{18}~cm^{-2}}$,
which is in a better agreement with the edge measurement. The remaining
small discrepancy is likely caused by a slight over-estimate of the
atomic oxygen column density from the edge measurement, 
presumably because of neglecting the 
compound contribution, as acknowledged by \citet{jue04}. Furthermore, their
edge model as plotted seems to overpredict the absorption at the edge 
($\lesssim 23$ \AA; see their Fig.~4).

\begin{figure}
  \centerline{
	\includegraphics[width=0.45\textwidth]{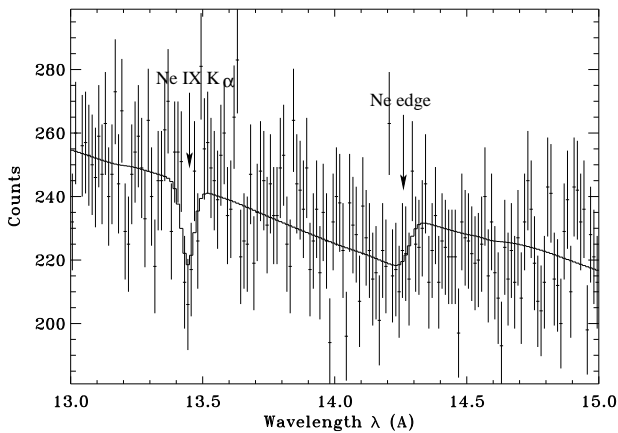}
  }
  \caption{The Ne edge and the \neix\ K$\alpha$ absorption line in the 
    HRC-LETG spectrum, fitted with XSPEC {\sl edge} plus {\sl absline} models 
    ($\chi^2/d.o.f. = 150/144$). The bin size is 0.0125 \AA; the ARF and RMF
    have been applied to the model. The rest is the same as in 
    Fig.~\ref{fig:OI}.
    \label{fig:NeEdge}
  }
\end{figure}

We can also measure the column density
of neon based on its absorption edge. Because neon is a noble element and 
is not easily in a compound form, the absorption edge structure is much 
simpler than that of oxygen. But, in the HETG observations, the neon edge 
of the MEG negative grating arm data unfortunately lands in a CCD gap, 
and there are also some complicated instrumental features around the edge 
wavelength of the positive arm data. We thus use only the LETG spectrum 
for the neon edge measurement.
We fit the spectrum in the wavelength range of 2-24.5 \AA\ 
with the XSPEC {\sl varabs} model, assuming the solar abundances, 
except for the neon,
which is modeled separately with 
the XSPEC {\sl edge} model (Fig.~\ref{fig:NeEdge}).
We obtain the neon edge wavelength 
$\lambda_E=$14.28(14.23, 14.35) \AA\ 
and optical depth $\tau_E=8.6(7.0, 10.2)\times10^{-2}$. 
Adopting the neon absorption cross section of 
$3.67\times10^{-19}$ cm$^{-2}$ from \citet{bal92}, 
we derive the neon
column density as $N{\rm_{Ne}=2.3(1.9, 2.7)\times10^{17}~cm^{-2}}$.

By combining these measurements of the oxygen and neon column
densities, we further estimate the Ne/O ratio in the cold plus warm gas
as (Ne/O)$_{\rm cw} = N_{\rm Ne}/N$(\ion{O}{1}+\ion{O}{2}+\ion{O}{3})
=0.3(0.2, 0.5), or 2.1(1.3, 3.5) solar.

\subsection{Hot Phase}

We first consider the X-ray absorption by hot gas of a single temperature and a
single velocity dispersion (neglecting the small ion mass differences
between oxygen and neon).
Adopting \ovii\ as the reference ion, Table~\ref{tab:absline}
illustrates, step by step, the dependence of 
parameter constraints on the inclusion of individual absorption lines.
The logic in the parameter constraints in each step is described in the following:

1. Fitting the \ovii~K$\alpha$ absorption line alone provides an upper 
limit to $b_v$ (determined by the instrument resolution)
and a corresponding lower bound to $N_{\rm OVII}$
(whereas its upper limit  corresponds to $b_v\simeq0$ km~s$^{-1}$). 

2. Jointly fitting the \ovii~K$\alpha$ and K$\beta$ lines gives
a lower limit to $b_v$ and therefore an improved upper limit to 
$N_{\rm OVII}$. Because both lines trace the same ion, an extremely 
small $b_v$ would make the \ovii~K$\alpha$ line too saturated to be 
consistent with the observed
\ovii~K$\beta$/K$\alpha$ constraint.

3. The inclusion of the \oviii\ line gives a measurement of 
$T$ (assuming a uniform isothermal CIE plasma origin for the lines). 
 
4. Finally, the \neix\ line, together with the measurement of $T$ in Step 3,
constrains the neon column density.
The comparison of the neon and oxygen column densities 
then leads to the estimate of the Ne/O ratio.

\begin{deluxetable*}{lcclcc}
  \tablewidth{0pt}
  \tablecaption{Step-by-step {\sl absline} fits to hot gas absorption
    lines \label{tab:absline}}
  \small
  \tablehead{ 
   Step &   Included line(s)                     & $b_v$       & log $N_{\rm OVII}$ & log $T$(K)  & \\
    &  & (km~s$^{-1}$)& (cm$^{-2}$)   &       & 
    Ne/O }
  \startdata
   1& \ovii~K$\alpha$ &  $<446$  & 17.2(16.3, 18.7)& $\cdots$ & $\cdots$ \\
   2 &  \ovii~K$\alpha$, K$\beta$  & 
    298(169, 505) & 16.3(16.1, 16.5) & $\cdots$ & $\cdots$ \\
  3 &  \ovii~K$\alpha$, K$\beta$, \oviii~K$\alpha$ &
    325(197, 490) & 16.3(16.1, 16.5) & 6.34(6.29, 6.41) & $\cdots$\\
  4 &  \ovii~K$\alpha$, K$\beta$, \oviii~K$\alpha$, \neix~K$\alpha$ &
    255(165, 369) & 16.3(16.1, 16.5)$^a$  & 6.34(6.29, 6.41) & 1.4(0.9, 2.1)$^b$
    \enddata
  \tablecomments{
    $^a$ The equivalent hydrogen, \oviii, and \neix~column densities 
    are log[$N_H$(cm$^{-2}$)]=20.1(19.9, 20.2),
    log[$N_{\rm OVIII}$(cm$^{-2}$)] = 16.4(16.2, 16.6), and
    log[$N_{\rm NeIX}$(cm$^{-2}$)] = 16.0(15.9, 16.1). 
    $^b$ The Ne/O ratio is in units of the AG89 solar 
    number ratio. }
\end{deluxetable*}

\begin{figure}
  \centerline{
	\includegraphics[width=0.45\textwidth]{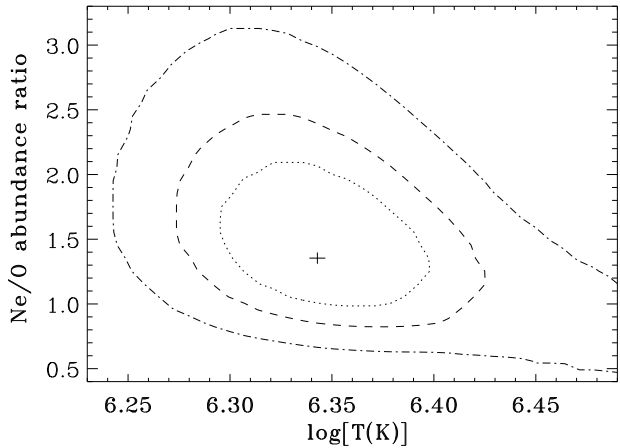}
  }
  \caption{The 60\%, 90\%, and 99\%
    confidence contours of the Ne/O ratio and gas temperature, 
   as obtained in the final joint-fit of the hot gas X-ray
   absorption lines with the {\sl absline} model 
   (Step 4 in Table~\ref{tab:absline}).
   \label{fig:contours} }
\end{figure}

The line centroid shifts obtained from the final joint-fit 
[Fig.~\ref{fig:OI} (b)-(d)] are nearly the same as those obtained
from the Gaussian model fits (Table~\ref{tab:gauss}).
The parameters ($T$, $b_v$, and $N_{\rm O{\small VII}}$), 
though more tightly constrained here, 
are still consistent with those found in  the previous studies within the 
quoted error ranges (Futamoto \etal 2004; Paper I). But the measurement of 
the Ne/O ratio here is completely new. Fig.~\ref{fig:contours} shows the 
confidence contours of the Ne/O ratio vs. $T$.

One may expect variation in physical and chemical properties of the hot gas 
along the sight line to \source. The so-called Local Bubble around the Sun, 
for example, is believed to contain hot gas with 
$T\sim10^6$ K (e.g., Kuntz \& Snowden 2000), lower than the average value
along the sight line (Table~\ref{tab:absline}). Because the Sun is not
supposed to be at a ``privileged'' spot in the Galaxy, it is natural to 
assume that much of the Galactic disk is occupied by such bubbles,
and that somewhat hotter versions of which may be responsible for the 
ubiquitous 3/4 keV 
background emission \citep{sno97}. 
X-ray absorption line spectroscopy of multiple sight lines has also shown
evidence for the presence of a thick Galactic disk with a scale height 
of 1-2 kpc \citep{yao05}. Furthermore, the Galactic bulge is apparently 
enhanced in X-ray emission (Snowden et al. 1997), although whether it 
arises mostly from unresolved point-like sources or 
truely diffuse hot gas is yet to be determined. A preliminary comparison 
of the \source\ sight line
with those toward LMC~X--3 ($l, b = 273^\circ.58, -32^\circ.08, D=50$ kpc;
Paper II) and 4U~1957+11 (V1408 Aql; $l, b = 51^\circ.31, -9^\circ.33, 
D=5.5$ kpc; Yao \etal in preparation) shows that the averaged temperature 
of the hot gas along the sight line through the bulge appears hotter by a 
factor of $\sim 1.5$.

\begin{figure}
 \centerline{
    \includegraphics[width=0.4\textwidth]{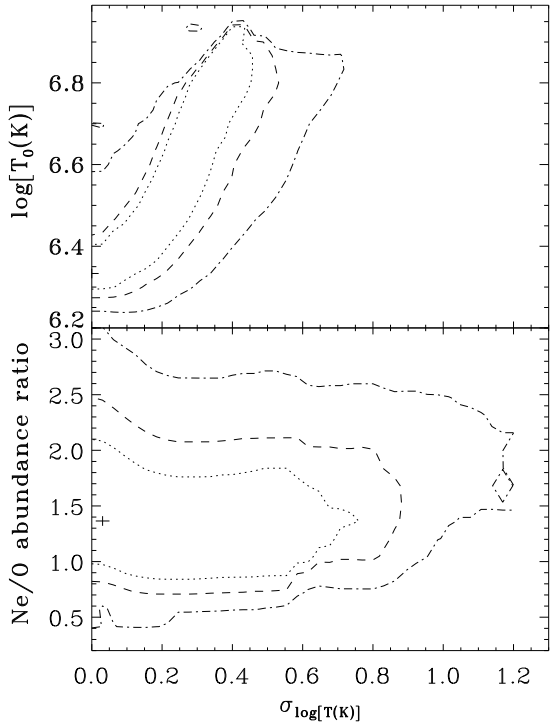}
   \vspace{0.05in}
 }
 \caption{The 68\%, 90\%, and 99\% confidence contours for the log-Gaussian temperature distribution. 
   \label{fig:gauss_T} }
\end{figure}

Though without a detailed knowledge about the hot gas variation, we can still 
check its effect on our absorption line measurements. 
Because the size of a bubble similar to the local one is small 
(dimension $\sim 10^2$ pc), its individual contribution (e.g., $N_{\rm OVII}$ 
$\lsim3\times10^{15}$  cm$^{-2}$; \citet{fang03}) is negligible,
compared to the observed total absorption toward \source\ 
(Table~\ref{tab:absline}). Therefore,  we can characterize
the variation within a bubble and/or from one bubble to another along a 
sight line by using a dispersion around a mean parameter (e.g., $T$). 
This dispersion may also include any Galaxy-wide variation. 
Here we test how the Ne/O ratio may 
be influenced if the isothermal assumption is relaxed by adopting a simple 
extension to a log-Gaussian temperature distribution:

\begin{equation}
{\rm d}N(T) \propto e^{\frac{-({\rm~log}T - {\rm~log}T_0)^2}{2(\sigma_{{\rm log}T})^2}} {\rm d}{\rm log}(T), 
\end{equation}
where $T_0$  is equivalent to $T$ in the isothermal case and is in
units of K,  while $\sigma_{logT}$ is the dispersion of log$T$. After 
revising our {\sl absline} model to incorporate this distribution, 
we repeat the final joint-fit again. This fit gives a 90\% upper limit 
of 0.84 to the new parameter $\sigma_{{\rm log} T}$. As shown in the upper 
panel of  Fig.~\ref{fig:gauss_T}, there is a 
strong correlation between log$T_0$ and $\sigma_{{\rm log} T}$;
i.e., the fit prefers a higher centroid temperature $T_0$ when 
the dispersion is broader. This correlation results from
the observed strong \oviii\ absorption line, which arises only
in a narrow and relatively high temperature range 
[log$T$(K)$ \sim 6.35-6.50$], 
compared to \ovii, which dominates over a much 
broader and lower temperature range (see Fig. 1 in Paper I).
The lower panel of  Fig.~\ref{fig:gauss_T} shows that the inferred 
Ne/O ratio is insensitive to $\sigma_{{\rm log} T}$. A similar conclusion
is also reached in our test with a power law temperature distribution.

\section{Oxygen Abundances}

In addition to the preceding X-ray oxygen column density 
and Ne/O ratio results, we can also estimate the absolute 
oxygen abundances (O/H). Such estimates, however,
cannot be made for all individual phases separately, due to
the lack of required measurements. For example, we cannot separately
estimate the Ne/O ratios in the cold and warm phases, because the
Ne edge data do not allow for individual measurements of the corresponding
neon column densities. 
Table~\ref{tab:abundance} summarizes the results 
for various ISM phases and/or their combinations. 
To estimate an absolute oxygen abundance, we need to know  the 
hydrogen column density in the same phase and in the same form.
Because \source\ is located above the \ion{H}{1} disk of the Galaxy, 
we can use the hydrogen
column density from the 21 cm emission line survey in the source field,
$N$(\ion{H}{1})$=1.5\times10^{21}$ cm$^{-2}$ \citep{dick90}
\footnote{We use this easily accessible data base, instead of the newer,
probably more reliable Leiden-Argentine-Bonn HI survey \citep{kal05}. 
The difference ($\sim15\%$) between the two survey values is much 
less than the statistical uncertainties in the derived parameters.
},
which may be compared to a total cold hydrogen column density 
$N$(\ion{H}{1})+2$N$(H$_2$)$\simeq1.9\times10^{21}$ cm$^{-2}$ inferred
from the optical extinction of the global cluster $E(B-V)\simeq0.32$
\citep{boh78, kuu03}. We estimate oxygen abundance in the cold phase as
$ ({\rm O/H})_n = N$(\ion{O}{1})$/N$(\ion{H}{1})$ = 2.6(1.3, 5.3)\times 10^{-4}$, or 
0.3(0.2, 0.6) AG89 solar.

\begin{deluxetable}{lccc}
  \tablewidth{0pt}
  \tablecaption{Summary of Results   \label{tab:abundance} }
  \tablehead{ 
    & \multicolumn{3}{c}{\underline{~~~~~~~~~~~~~~~~~~~~Gas Phase~~~~~~~~~~~~~~~~~~~~}} \\ 
    Parameter    & cold & warm & hot }
  \startdata
         & \multicolumn{3}{c}{column density} \\
     O   & 17.6(17.3, 17.9) & 17.5(17.2, 17.8)$^a$ & 16.7(16.5, 16.8)$^b$\\ 
         & \multicolumn{3}{c}{17.9(17.7, 18.1)$^c$~~~17.6(17.3, 17.8)$^d$}\\
     H   & 21.2$^e$ & \multicolumn{2}{c}{20.4$^{d,f}$} \\  
     Ne  & \multicolumn{2}{c}{17.4(17.3, 17.5)$^c$} & 16.0(15.9, 16.1) \\ 
     \hline
         & \multicolumn{3}{c}{Abundance$^g$} \\ 
    O/H  & 0.3(0.2, 0.6) & 2.0(0.8, 3.6) & $\gsim0.94$ \\
         & \multicolumn{3}{c}{0.5(0.3, 0.9)$^c$~~~~~~~2.2(1.1, 3.5)$^d$} \\
    Ne/O & \multicolumn{2}{c}{2.1(1.3, 3.5)$^c$} & 1.4(0.9, 2.1)
    \enddata
    \tablecomments{
      The column densities are all on a logarithmic scale. The values for
      the cold and warm oxygen phases are
      obtained from the fits to individual lines (\oi\ K$\alpha$, 
      \oii\ K$\alpha$, and \oiii\ K$\alpha$), whereas the values for the hot 
      phase are obtained from a joint-fit to multiple lines 
      (\ovii\ K$\alpha$, K$\beta$, \oviii\ K$\alpha$, and \neix\ K$\alpha$).
      $^a$ N(\ion{O}{2}+\ion{O}{3}), which may be over-estimated because
of the contamination from a potential compound oxygen line that is not 
accounted for.
      $^b$ N(OVII+OVIII).
      $^c$ Cold plus warm gas.
      $^d$ Warm plus hot (ionized) gas.      
      $^e$ The 21 cm emission data are used.
      $^f$ The pulsar dispersion measure is used.
      $^g$ All abundances and ratios are listed in units of the AG89 
      solar values
      (i.e., O/H = 8.5$\times10^{-4}$ and Ne/O = 0.144). }
\label{tab:summary}
\end{deluxetable}

For ionized gas, we can use the free electron column density $N_e$
from the pulsar DM (\S 2). 
The oxygen column densities in the warm and hot 
phases are $N$(\ion{O}{2}+\ion{O}{3}) and $N$(OVII+OVIII), neglecting  
\oiv\ - \ovi\ and \oix, which are only important in thermally
very unstable ``intermediate'' temperatures ($T\lsim10^{5.5}$ K)
and in probably rare low-density regions with temperatures 
$\gtrsim 10^{6.5}$ K. We can then define the oxygen abundances in
the two phases as $({\rm O/H})_w = N$(\ion{O}{2}+\ion{O}{3})/$\eta (1-\xi) N_e$ and 
$({\rm O/H})_h = N($\ion{O}{7}+\ion{O}{8}$)/\xi \eta N_e$,
where $\xi$ is  the hot phase fraction  of the 
electrons and $\eta = 0.84$ accounts for the contribution from helium.
Letting $({\rm O/H})_h = \alpha$ (O/H)$_w$, we have $\xi = 1/(1+\alpha r_N$),
where $r_N = N($\ion{O}{2}+\ion{O}{3}$)/N($\ion{O}{7}+\ion{O}{8}$) \sim 8$.
We expect that dust grain destruction and chemical enrichment occurs 
primarily in the hot phase; i.e., the metal abundance in the hot phase
should be comparable to, or higher than, that in the warm phase. 
Therefore, $\alpha \gtrsim 1$; specifically, 
$({\rm O/H})_w = 1.8(0.8, 3.1)\times10^{-3}$ for $\alpha =1$ and
$1.6(0.7,2.8) \times10^{-3}$ for $\alpha = \infty$.
We thus adopt 1.7(0.7, 3.1)$\times10^{-3}$ 
[or 2.0(0.8, 3.6) solar] to account for
this small $\alpha$-dependent uncertainty. Systematically,  however, 
this oxygen abundance could be
an overestimate, because of the potential contamination to the \ion{O}{2} line 
by the uncertain compound oxygen line as mentioned in \S~3.
For instance, if the compound oxygen column density is 
$\sim0.5N{\rm _{OI}}$, as suggested by \citet{tak02}, then 
$({\rm O/H})_w \sim 9.2\times10^{-4}$ and
$6.7\times10^{-4}$ for $\alpha =1$ and $\infty$, respectively.
On the other hand, $({\rm O/H})_h \propto \alpha$
cannot be determined without knowing $\alpha$. Nevertheless, we 
may estimate the mean oxygen abundance in total ionized gas:
[$N({\rm O{\small II}+O{\small III}})$ + $N({\rm O{\small VII}+O{\small VIII}})$]/$\eta N_e = 2.2(1.1, 3.5)$ solar, or $\sim1.1$ solar if the potential
contamination from the compound oxygen is accounted for.


\section{Comparisons with Other Abundance Measurements}

Not all X-ray binaries are suitable background sources
for ISM measurements. High mass X-ray binaries (HMXBs)
suffer potentially serious stellar contamination 
to absorption features. The observed X-ray-absorbing column density 
of the HMXB Cyg X--1, for example, shows an orbital dependence 
\citep{ebi96, sch02b}. Even LMXBs could be problematic.
There are previous Ne/O measurements based on absorption edge data
of several LMXBs \citep{pae01, jue03, jue05}. Higher Ne/O 
ratios (up to a factor of several times the solar value) are sometimes reported
and are suggested to be local to the binary systems, which typically show
burst behaviors and are 
proposed to have  neon-rich white dwarf companions. The 
discovery of the Ne/O ratio variation with time in two of such sources 
supports this scenario \citep{jue05, sid05}. 

Relatively secure X-ray estimates of the ISM oxygen and neon abundances
in previous studies are available only for the sight line to Cyg X--2 
\citep{tak02}. With its Galactic position 
($l, b = 87^\circ.33, -11^\circ.32$ and distance = 7.2 kpc),
the sight line samples gas primarily at the solar radius of the Galaxy, 
providing an interesting comparison
with the inner Galactic disk sight line to \source. Similar to \source,
Cyg X--2 is located at a large off-Galactic plane distance, allowing 
reasonable estimates of the 
hydrogen column densities along the sight line from
21 cm, H$\alpha$, and CO emission lines. \citet{tak02} estimated that the
oxygen abundance  is $\sim 0.47 \pm 0.16$ solar in the atomic gas and
a factor of $\sim 1.5$ higher in the atomic plus compound ISM. 
Their estimates are based on the absorption edges, which are dominated by
the cold and warm phases. 
Although our oxygen  measurements are based on the absorption
lines, we can derive an average oxygen abundance for the cold plus warm
gas along the sight line to \source\ 
(Table~\ref{tab:abundance}), 
$N$(\ion{O}{1}+\ion{O}{2}+\ion{O}{3})/[$N$(HI)+$(1-\xi)\eta N_e$]=0.52(0.33, 0.85) solar.
This value is consistent with the oxygen 
abundance estimate along the sight line to Cyg X--2. 

Our neon abundance measurement of the cold plus warm gas 
along the sight line to
\source\ is based on the absorption edge, as it is to Cyg X--2 
\citep{tak02}. The neon abundance 0.75(0.55, 0.95) toward this latter source
is considerably lower than 1.2(1.0, 1.4) inferred here for the sight
line to \source, which indicates a moderately metal enhancement
in the inner region compared to the outer region of the Galaxy. 
The Ne/O ratios along these 
two sight lines are compatible [2.1(1.3, 3.5) vs. 1.6(0.9, 2.3)]. 
For \source, we also have
the hot Ne/O ratio 1.4(0.9, 2.1), 
comparable to the value in the cold plus warm gas, particularly 
if the expected oxygen compound contribution is  included.
The hot Ne/O ratio, if typical in the Galaxy, 
can be used in a joint analysis of oxygen and neon X-ray absorption lines
observed in other sight lines. 

Now let us compare our inferred ISM Ne/O 
ratios with those stellar measurements. At first glance,  our Ne/O 
ratio [$\sim 2.1(1.3, 3.5)$] obtained for the cold plus warm gas
is consistent with the high Ne/O ratio as suggested by \citet{drake05}. 
But, including as well the expected oxygen compound contribution 
would then lead to a reduced Ne/O ratio 
(by a factor of $\sim 1.5$; Takei \etal 2002) that
would be significantly lower than the stellar ratio. Similarly, 
our inferred hot gas Ne/O ratio  (Table~\ref{tab:summary}) is significantly
lower (by $\sim3\ \sigma$) than the stellar value.
It is difficult to reconcile this Ne/O ratio difference and the total oxygen 
abundance similarity between the stellar and ISM measurements. There is
no obvious way to hide neon in the ISM! 
We suspect that the Ne/O ratio estimate from stellar 
emission lines, which is generally sensitive to modeling assumptions, 
may not represent the true underlying composition of the stars.
The inconsistency of the stellar measurements by \citet{drake05}
with the recent solar studies \citep{sch05, young05} illustrates
such an uncertainty.

\section{Summary\label{sec:summary}}

We have used the \chandra HETG and LETG observations 
of the LMXB \source\ as a test case to explore the potential for
 metal abundance measurement  in different atomic phases of the ISM. We have 
concentrated on the measurements of atomic oxygen and neon, as summarized in
Table~\ref{tab:summary}, and 
on comparison with previous X-ray studies.
Our main results and conclusions are as follows:

1. We have separately measured the column densities of 
\oi, \oii, and \oiii\ from their heavily saturated K$\alpha$ lines. 
These
measurements are only weakly dependent on the exact velocity dispersion
assumed. 
A comparison
of the \oi\ column density and the 21 cm hydrogen emission in the field
gives a cold (neutral) oxygen abundance of 0.3(0.2, 0.6) solar (AG89).
 The ratio of our measured \oii\ plus \oiii~column density to the 
pulsar DM along the same sight line further gives an estimate
of the warm oxygen abundance as 2.0(0.8, 3.6)  solar.

2. We have constrained the neon column density from 
its absorption edge, giving an  abundance 
of 1.2(1.0, 1.5) solar and a Ne/O ratio of 2.1(1.3, 3.5) solar for 
the cold plus warm gas. The Ne/O ratio with the inclusion of the compound 
ISM is likely to be a factor of $\sim 1.5$ lower \citep{tak02}. 

3. We have also measured the oxygen column density and Ne/O ratio 
in the hot ISM,
based on a joint-analysis of the detected \ovii~K$\alpha$, \oviii~K$\alpha$, 
and \neix~K$\alpha$ absorption lines, together with the non-detection
of the \ovii~K$\beta$ line.
Assuming that \ovii~K$\alpha$ and \oviii~K$\alpha$
trace all hot gas (i.e., $T \sim 10^{5.5-6.5}$ K), we estimate
that the hot phase accounts for about 6\% of
the total oxygen column density along the sight line and obtain a 
Ne/O ratio of 1.4(0.9, 2.1)  solar,
which is insensitive to the exact temperature distribution assumed. 

4. Our abundance estimates for the atomic phases, 
together with complementary X-ray spectroscopic studies of the total
abundances, may have strong implications for understanding various
chemical enrichment and depletion processes both in the ISM and during
star formation. There is evidence for an enhanced enrichment
in the hot gas, especially in comparison with the cold phase.
The Ne/O ratios obtained in the ISM are significantly 
smaller than the value indicated in the recent emission line 
measurement of solar-like stars \citep{drake05}.

The existing X-ray measurements of the ISM abundances 
(including the present work) are still preliminary, subject to various 
systematic uncertainties, both theoretical and observational, which are 
yet difficult to fully quantify. Nevertheless, the outcome of this
study demonstrates the unique potential for a 
comprehensive characterization of the metal abundances in various ISM forms
and phases. 

\acknowledgments
We thank Daniel Dewey, Edward Jenkins, and Blair Savage for reading an
early draft of the paper and for providing valuable comments and suggestions.
We also thank the referee for valuable suggestions which helped to 
improve the presentation of the paper.
We are grateful to Todd Tripp, Paola Testa, and Norbert Schulz 
for useful discussions. 
Support for this work was partially provided by NASA through the 
Smithsonian Astrophysical Observatory (SAO) contract SV3-73016 to the 
MIT for support of the Chandra X-Ray Center, which is operated by the 
SAO for and on behalf of NASA under contract NAS 8-03060. 
Support from a {\sl Chandra} archival research grant is acknowledged.

\end{document}